\documentclass[prb,aps,twocolumn,superscriptaddress,showpacs]{revtex4}
\usepackage{amsfonts,amsmath,bm}
\usepackage[T1]{fontenc}
\usepackage{color}
\usepackage[pdftex]{graphicx}
\usepackage{epstopdf}

\definecolor{gray}{rgb}{0.45,0.45,0.45}

\newcommand{\rr}{{\bm{r}}}
\newcommand{\kk}{\bm{k}}

\newcommand{\kp}{\bm{k}\!\cdot\!\bm{p}}

\newcommand{\Leps}{\text{\Large{$ \varepsilon $}}}

\begin{document}

\title{Electron states in a double quantum dot with broken axial symmetry} 

\author{Krzysztof Gawarecki}
\email{Krzysztof.Gawarecki@pwr.wroc.pl} 
\affiliation{Institute of Physics, Wroc{\l}aw University of
Technology, 50-370 Wroc{\l}aw, Poland}

\author{Pawe{\l} Machnikowski}
\affiliation{Institute of Physics, Wroc{\l}aw University of
Technology, 50-370 Wroc{\l}aw, Poland}
\author{Tilmann Kuhn}
\affiliation{Institut f\"ur Festk\"orpertheorie, Universit\"at
  M\"unster, 48149 M\"unster, Germany}

\begin{abstract}
We study theoretically the electron states in a system of two vertically stacked quantum dots. We investigate the influence of the geometrical symmetry breaking (caused by the displacement as well as the ellipticity of the dots) on the electron states. 
Our modeling is based on the $8$-band  $\bm{k}\!\cdot\!\bm{p}$ method. We show that the absence of axial symmetry of the system leads to a coupling of the $s$ state from one dot with the $p$ and $d$ states from the other. Our findings indicate, that this coupling can produce a strong energy splitting at resonance (on the order of several meV) in the case of closely spaced quantum dots. Furthermore, we show that in the presence of a piezoelectric field, the direction of the displacement plays an important role in the character of the coupling. 
\end{abstract}


\maketitle

\section{Introduction} 
\label{sec:intro}

Systems composed of vertically stacked double quantum dots (DQDs) show many interesting properties.\cite{wu14} Furthermore, DQDs have been proposed to be used for quantum-coherent devices, including spin-based quantum bits.\cite{doty10} Pairs of self-assembled quantum dots (QDs), are particularly interesting because they can be relatively easily produced in a Stransky-Krastanov process. Because in such systems the dots are placed rather close to each other, their properties are significantly affected by tunnel coupling \cite{bryant93,bayer01,korkusinski01,schliwa01,szafran01,bester04,gawarecki10,gawarecki11a,gawarecki12a}. 

Carrier spectra of quantum dots have been widely described in the literature using the $\kp$ model\cite{korkusinski01,schliwa01,schliwa07,stier99,andrzejewski10,voon2009k} as well as tight-binding and pseudopotenial methods \cite{bester05,bester06a,williamson00,sundaresan10,zielinski12,zielinski13,usman11,usman12}.  However, DQDs composed of lens shaped QDs, are often modeled  assuming the axial symmetry of the system. In that approximation, the axial projection of the envelope angular momentum is conserved and there is no coupling between states with different angular momenta. On the other hand, from the experimental point of view, samples usually do not have axial symmetry\cite{wang02} (dots from different layers can be shifted with respect to each other and can be elliptical). This opens the possibility of an additional coupling, which would be prohibited in an ideal (symmetric) case. Indeed, some experiments exhibit features, which suggest such a behavior \cite{mueller12}. In Ref. \onlinecite{daniels13} the symmetry breaking in a DQD excitonic system was studied. However the calculations have been performed in the effective mass approximation and the deviations from the axial symmetry (due to a displacement and an ellipticity) have been introduced by a small perturbative parameter. 

In this work, we study systematically the influence of the geometrical axial symmetry breaking of arbitrary magnitude on the electron states in the structure composed of two vertically stacked QDs formed in
the Stransky--Krastanov self-assembly process. We consider $\mathrm{In_{0.8}Ga_{0.2}As}$ dots embedded in a GaAs matrix.
We calculate the strain distribution in the system using the continuous elasticity approach\cite{pryor98b}. The piezoelectric field is included up to second order in the strain tensor \cite{bester06b,bester06a} which leads to a dependence of the predicted spectral features on the direction of the system deformation with respect to the crystallographic axes. We find the electron states within the $8$-band $\kp$ model. We show that axial symmetry breaking in a DQD structure leads to a qualitative reconstruction of the energy spectrum, in particular in the vicinity of level crossings. This effect turns out to depend crucially on the system geometry with respect to the crystallographic axes. 

The paper is organized as follows. In Sec.~\ref{sec:model}, we define
the model. In Sec.~\ref{sec:results}, we
discuss results of the obtained electron states. Finally, concluding remarks and
discussion are contained in Sec.~\ref{sec:concl}.

\section{Model} 
\label{sec:model}

The system under consideration contains two vertically stacked $\mathrm{In_{0.8}\mathrm{Ga}_{0.2}As}$ QDs, where we assume a homogeneous alloying.
Both dots are placed on wetting layers (with an assumed width of $0.6$~nm). 
Because of a lattice mismatch between InAs and GaAs, strain appears in the system. In order to find the strain distribution we performed a minimization of the elastic energy of the system\cite{pryor98b} using the continuous elasticity approach. As a result, we obtained the displacement field and the strain tensor $\epsilon$.

Due to a non-zero shear strain in the system, a piezoelectric (PZ) field appears and affects the carrier states\cite{bester06a}. In order to calculate the potential generated by the piezoelectricity ($V_{\mathrm{PZ}}$), we calculated the polarization of the system up to second order in the strain tensor. A detailed description of the piezoelectric field calculation is given in Appendix \ref{sec:app1}.
The local band structure is derived from the 8-band $\kp$
Hamiltonian with the strain-induced terms. Because of its numerical advantages\cite{andrzejewski13} we use the LS basis \{$|S\uparrow\rangle,|X\uparrow\rangle,|Y\uparrow\rangle,|Z\uparrow\rangle,|S\downarrow\rangle,|X\downarrow\rangle,|Y\downarrow\rangle,|Z\downarrow\rangle$\} where $S,X,Y,Z$ denote electron orbitals and $\uparrow$ and $\downarrow$ represent spin projection. In the matrix representation the Hamiltonian takes the form\cite{schliwap}
\begin{equation*}
\label{matrix2x2}
H =
\left (  
\begin{array}{cc}
H(\kk) & \Gamma \\
-\Gamma^{*} & H(\kk) \\
\end{array}
\right ),
\end{equation*}
where
\begin{equation*}
\Gamma = i \frac{\Delta}{3}
\left (  
\begin{array}{cccc}
0& 0 & 0 & 0 \\
0& 0 & 0 & 1 \\
0& 0 & 0 & -1 \\
0& -1 & 1 & 0 \\
\end{array}
\right )
\end{equation*}
and $H(\kk) = H_{1}+H_{2}$. Here
\begin{equation*}
H_{1} =
\left (  
\begin{array}{cccc}
E_{s} & i P k_{x} &  i P k_{y}  &  i P k_{z} \\
-i P k_{x}& E_{x} & N'k_{x}k_{y}-i\frac{\Delta}{3} & N'k_{x}k_{z} \\
-i P k_{y}& N'k_{x}k_{y}+i\frac{\Delta}{3} & E_{y} & N'k_{y}k_{z} \\
-i P k_{z}& N'k_{x}k_{z} & N'k_{y}k_{z} & E_{z} \\
\end{array}
\right )
\end{equation*}
and
\begin{equation*}
H_{2} =
\left (  
\begin{matrix}
0 & -i P \epsilon_{xj} k_{j} &  -i P \epsilon_{yj} k_{j}  &  -i P  \epsilon_{zj} k_{j} \\
i P \epsilon_{xj} k_{j}& 0 & n \epsilon_{xy} & n \epsilon_{xz} \\
i P \epsilon_{yj} k_{j}& n \epsilon_{xy} & 0 & n \epsilon_{yz} \\
i P \epsilon_{zj} k_{j}& n \epsilon_{xz} & n \epsilon_{yz} & 0 \\
\end{matrix}
\right ).
\end{equation*}
The diagonal part of $H_{1}$ contains 
\begin{align*}
E_{\mathrm{s}}  & =  A' (k^{2}_{x}+k^{2}_{y}+k^{2}_{z}) + E_{\mathrm{c}} + a_{\mathrm{c}}(\epsilon_{xx}+\epsilon_{yy}+\epsilon_{zz}), \\
E_{\mathrm{x}}  & =  L' k^{2}_{x} + M' (k^{2}_{y} + k^{2}_{z}) + E'_{\mathrm{v}} + l \epsilon_{xx} + m (\epsilon_{yy}+\epsilon_{zz}), \\
E_{\mathrm{y}}  & =  L' k^{2}_{y} + M' (k^{2}_{x} + k^{2}_{z}) + E'_{\mathrm{v}} + l \epsilon_{yy} + m (\epsilon_{xx}+\epsilon_{zz}), \\
E_{\mathrm{z}}  & =  L' k^{2}_{z} + M' (k^{2}_{x} + k^{2}_{y}) + E'_{\mathrm{v}} + l \epsilon_{zz} + m (\epsilon_{yy}+\epsilon_{xx}),
\end{align*}
with
\begin{align*}
E_{\mathrm{c}}  & =  E_{\mathrm{v}} + E_{g} + V_{\mathrm{PZ}} -e \Leps z,\\
E'_{\mathrm{v}}  & =  E_{\mathrm{v}} - \Delta/3 + V_{\mathrm{PZ}} -e \Leps z,\\
A' &= \frac{\hbar^{2}}{2m_{0}} \left ( \frac{1}{m^{*}_{e}} - \frac{E_{\mathrm{p}}}{E_{g}} \frac{E_{\mathrm{p}}(E_{g}+2\Delta/3)}{E_{g}(E_{g}+\Delta)} \right ), \\
E_{\mathrm{p}}&=\frac{2m_{0} P^{2}}{\hbar^{2}},\\
L' & = \frac{P^{2}}{E_{g}}-\frac{\hbar^{2}}{2m_{0}} (\gamma_{1}+4 \gamma_{2}),\\
M'  & =  -\frac{\hbar^{2}}{2m_{0}} (\gamma_{1} -2 \gamma_{2}),\\
N' &= \frac{P^{2}}{E_{g}}-\frac{3 \hbar^{2}}{m_{0}} (\gamma_{1}+4 \gamma_{2}),
\end{align*}
where $E_{\mathrm{v}}$ denotes the unstrained average valence band edge, $\Delta$ is the spin-orbit split-off element, $E_{\mathrm{g}}$ is the energy gap, $P$ is a parameter proportional to the interband momentum matrix element,  $m_{0}$ is the free electron mass, $m^{*}_{e}$ is the electron effective mass in a bulk material, $\Leps$ denotes the axial electric field and $\gamma_{i}$ are Luttinger parameters. In $H_{2}$ the Einstein summation convention is being used.
The influence of the strain field on the carrier states has been accounted for using
$l = 2 b_{\mathrm{v}} + a_{\mathrm{v}}$,
$m = a_{\mathrm{v}} - b_{\mathrm{v}}$,
$n = \sqrt{3} d_{\mathrm{v}}$,
where $a_{\mathrm{c}},a_{\mathrm{v}},b_{\mathrm{v}}$ are the conduction and valence band deformation
potentials and $d_{\mathrm{v}}$ is the shear strain deformation potential. We perform Burt-Foreman ordering\cite{burt92,foreman97}, which for the upper triangular matrix is
$N' k_{i} k_{j} \rightarrow k_{i} N_{+} k_{j} + k_{j} N_{-} k_{i}$ and for the lower one $N' k_{i} k_{j} \rightarrow k_{j} N_{+} k_{i} + k_{i} N_{-} k_{j}$ where $N_{-}=M'-\hbar^{2}/2m_{0}$ and  
$N_{+}=N'- N_{-}$. In order to avoid spurious solutions we use the reduced value of $E_{\mathrm{p}}$\cite{xian10}.

Spin-orbit coupling (the Dresselhaus term) in the conduction band is neglected. The values of the material parameters are given in Table~\ref{tab:param}.
\begin{table}
\begin{tabular}{llll}
\hline
& GaAs & InAs & Interpolation of $\mathrm{In_{x}Ga_{1-x}As}$\\
\hline
 $E_{\mathrm{v}0}$ & 0.0~eV & 0.173~eV & 0.173x+0.058x(1-x)\\
 $E_{\mathrm{g}}$ & 1.518~eV & 0.413~eV & 0.413x+1.518(1-x)-0.477x(1-x)\\ 
 $E_{\mathrm{p}}$ & 21.0~eV & 18.0~eV & 18.0x+21.0(1-x)+1.48x(1-x)\\
 $m^{*}_{\mathrm{e}}$ & 0.065 & 0.022 & 0.022x + 0.065(1-x)-0.0091x(1-x)\\    
 $\Delta$ & 0.34~eV & 0.38~eV& 0.38x+0.34(1-x)-0.15x(1-x) \\ 
 $a_{\mathrm{c}}$ & -7.17~eV & -5.08~eV & -5.08x-7.17(1-x)-2.61x(1-x)\\
 $a_{\mathrm{v}}$ & 1.16~eV & 1.0~eV & linear\\
 $b_{\mathrm{v}}$ & -1.824~eV & -1.8~eV & linear\\
 $d_{\mathrm{v}}$ & -5.062~eV & -3.6~eV & linear\\ 
 $\gamma_{\mathrm{1}}$ & 19.7 & 7.1 & linear\\
 $\gamma_{\mathrm{2}}$ & 8.4 & 2.02 & linear\\
 $\gamma_{\mathrm{3}}$ & 9.3 & 2.91 & linear\\  
 $e_{14}$ & 0.230~$\mathrm{C/m^{2}}$ & 0.115~$\mathrm{C/m^{2}}$  & linear\\
 $B_{114}$ & -0.439~$\mathrm{C/m^{2}}$ & -0.531~$\mathrm{C/m^{2}}$  & linear\\
 $B_{124}$ & -3.765~$\mathrm{C/m^{2}}$ & -4.076~$\mathrm{C/m^{2}}$ & linear\\
 $B_{156}$ & -0.492~$\mathrm{C/m^{2}}$ &-0.120~$\mathrm{C/m^{2}}$  & linear \\
\hline
\end{tabular}
\caption{\label{tab:param}Material parameters used in the calculations.\cite{vurgaftman01,schliwa07}}
\end{table}
%
The resulting eigenproblem is solved using the Jacobi-Davidson method.  All details of the calculations have been described in Appendix \ref{sec:app2}.
Finally, the in-plane probability density of $i$-th state is calculated according to
\begin{equation*}
\rho_{i}(x,y) = \sum_{m=1}^{8} \int_{-\infty}^{\infty} \psi_{i,m}^{*}(x,y,z) \psi_{i,m}(x,y,z) dz,
\end{equation*}
where $\psi_{i,m}(x,y,z)$ is the $m$-th component (subband) of the $i$-th eigenfunction.
\section{Results} 
\label{sec:results}

In this section we discuss the results of our calculations performed for a single QD as well as for a DQD.

\begin{figure}[tb]
\begin{center}
\includegraphics[width=85mm]{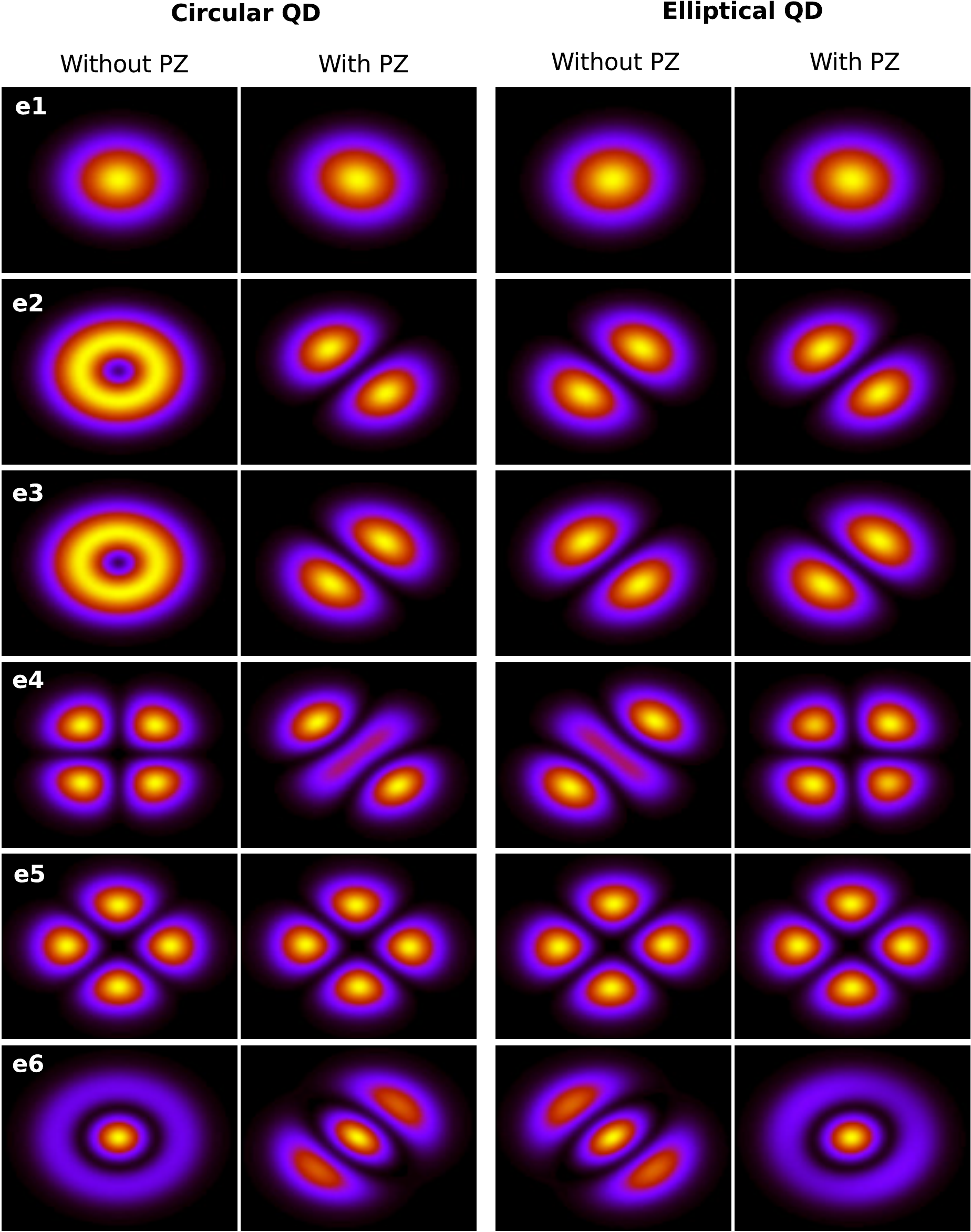}
\end{center}
\caption{\label{fig:smap}The in-plane probability density of the six lowest electron states. The first column corresponds to a circular lens-shaped QD in an ideal case (without a PZ field). The second one shows the same QD but in the presence of the PZ field. The third column presents the results for an elliptical QD without a PZ field. The last column contains results for an elliptical QD with the PZ field.  } 
\end{figure}
First, in order to provide a clear interpretation of the further results for a DQD system, we calculated the electron states in a single QD. Each column of Fig.~\ref{fig:smap} presents the in-plane probability density of the six lowest electron states (e0-e5). We consider four cases: a circular (i.e. axially symmetric) lens-shaped QD with and without the PZ field, as well as an elliptical QD with and without the PZ field.  The first column corresponds to the ideal case (circular lens-shaped QD without the PZ field). At this point, our results reflect the well known properties of a single QD~\cite{schliwa07,bester06a}. The ground state  (e0) has a $s$-type symmetry. Since the system has the axial symmetry, the projection of the envelope angular momentum $M$ is a good quantum number, and the ground state corresponds to $n=0$ and $M=0$, where $n$ denotes the excitation of the radial part of the wavefunction. The next two states (e1,e2) show $p$-type symmetry (that is $n=0$ and $M=-1,1$). Subsequently, e3,e4 and e5 exhibit $d$ character. The states e3 and e4 correspond to the degenerate states with $n=0$ and $M=-2,2$. Due to numerical reasons (the discretization on a rectangluar grid) the degeneracy is slghtly lifted and two linear combinations of these states appear which are rotated with respect to each other by $45^{\circ}$.
 In the case of e5 we have clearly $n=1$ with $M=0$. In the presence of the piezoelectric field (second column of Fig.~\ref{fig:smap}) the character of the states is different. Due to the piezoelectric field the symmetry of the system is lowered from $C_{\infty}$ to $C_{2v}$ \cite{bester06a}. In that case, $M$ is no longer a good quantum number. The contribution from the second order term of the piezoelectric field has an opposite sign to the first order term and is very important \cite{bester06a,bester06b}. However in the case of an alloy, the 2nd order contribution is lowered due to its dependence on the hydrostatic strain which vanishes with increasing Ga admixture\cite{usman11}. Now, the direction along the lower values of the PZ field is favored. The $p$ states are clearly combined into orbitals $p_{1}\sim \sin{(\varphi-\pi/4)}$ and $p_{2}\sim \cos{(\varphi-\pi/4)}$ which have mutually perpendicular orientation. Furthermore, the character of the $d$ states is significantly changed. The e3 and e5 states couple and change their symmetry. The e4 state, which is compatible with the symmetry of the PZ field, remains uncoupled.  In the third and the fourth column of Fig.~\ref{fig:smap} the results for an elliptical QD (with the major to minor axis ratio of $1.1$) elongated in the ($110$) direction are shown. In that case, even without a PZ field, the axial symmetry is broken and the states which are elongated in the direction of the major axis are lowered in the energy compared to the states elongated in the direction of the minor axis. In the case of ellipticity ratio of $1.1$, adding the PZ field, the orientation of the $p$ states as in the case of the circular QD with PZ field is restored. Furthermore, for the $d$ states the PZ field
essentially compensates the elliptical anisotropy such that the spatial profiles of the circular QD without PZ field are recovered.
In order to check the importance of the valence band to conduction band coupling in the $\bm{k}\!\cdot\!\bm{p}$ Hamiltonian
we compared our results with those obtained from a single band effective mass calculation based on the L\"owdin elimination method
\cite{lowdin51}. The difference between the relative shell energy levels in both cases is up to 24\%.
\begin{figure}[tb]
\begin{center}
\includegraphics[width=85mm]{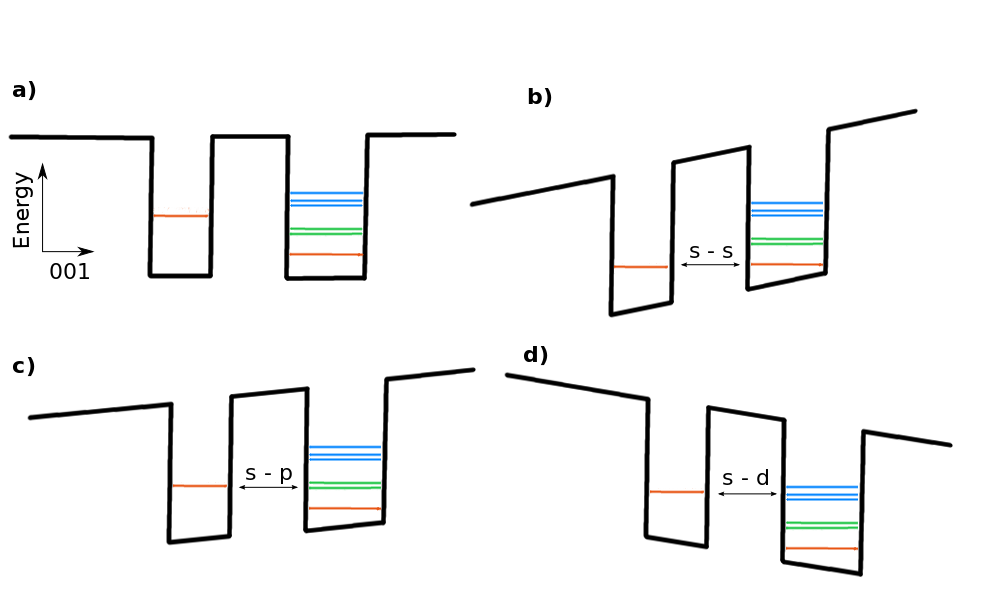}
\end{center}
\caption{\label{fig:schema}(Color online) Schematic electron energy structure in the investigated DQD without (a) and with different values of an electric field (b-d) indicating resonances between energy levels in the two dots.} 
\end{figure}
\begin{figure}[tb]
\begin{center}
\includegraphics[width=85mm]{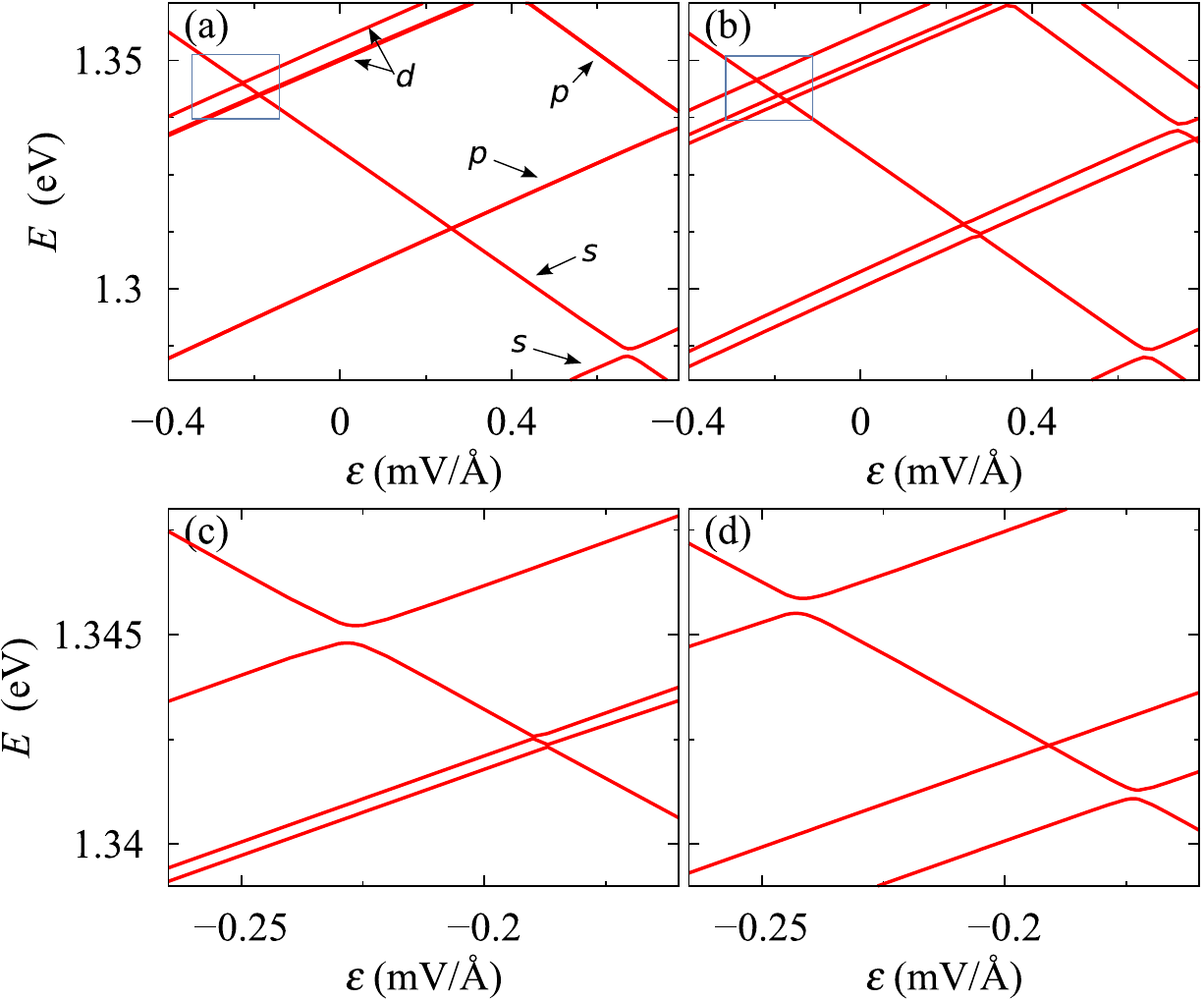}
\end{center}
\caption{\label{fig:general}(Color online) (a) Lowest electron energy branches as a function of the electric field without a PZ field. (b) The same as (a) but with the piezoelectric field included. (c) Enlarged region of (a) with $s$-$d$ resonances (indicated by the blue rectangle). (d) Enlarged region of (b) with $s$-$d$ resonances. } 
\end{figure}

Next, let us consider a DQD system with a geometrical axial symmetry. A schematic diagram which illustrates the energy structure in a DQD (where the dots have different sizes) is shown in Fig.~\ref{fig:schema}. The electronic structure can be tuned by applying an axial electric field which modifies the slopes of the band edges. From the experimental point of view, in a DQD system, the upper dot is often bigger than the lower one~\cite{krenner05}. Furthermore, in order to have $s$-$p$ and $s$-$d$ resonances at a reasonable value of the electric field, we assumed $r_{1}=9$~nm, $h_{1}=3.3$~nm and $r_{2}=12.6$~nm, $h_{2}=5.4$~nm, where $r_{1},h_{1}$ and $r_{2},h_{2}$ are base radius and height of the lower and the upper dot respectively\cite{gawarecki10}.

The electron energy levels for a fixed distance $D=10.2$~nm between the dots (counted from the base of the lower dot to the base of the upper one) are shown in Fig.~\ref{fig:general}(a,b). The dots are placed along the same $z$ ($001$) axis and the energy branches are shown as a function of the axial electric field. Fig.~\ref{fig:general}(a) presents the results without the piezoelectric field. At $\Leps=0.72$~mV/\AA~the electron $s$ states in both dots have similar energy (as shown in Fig.~\ref{fig:schema}b). Because the symmetry of these states allows them to couple, the energies show an anticrossing. At $\Leps=0.296$~mV/\AA~$s$ and $p$ states become degenerate (as shown in Fig.~\ref{fig:schema}c). In this case, the $p$ states are degenerate and there is no coupling between them and the $s$ type state from the second dot. In consequence, there is a crossing between the energy branches. 

The energy branches in the presence of the piezoelectric field are shown in Fig.~\ref{fig:general}(b). Because of the symmetry reduction due to the piezoelectric field, the electron states of type $p$ and $d$ are no longer degenerate. The splitting due to the PZ field is larger in the case of $p$ states than $d$ states, which is consistent with Ref.~\onlinecite{usman11}. However, in contrary to Ref.~\onlinecite{usman11}, we do not observe mixing between $s$ and $p$ states due to the PZ field. Figs.~\ref{fig:general}(c,d) present an enlarged part (marked by the blue box) of Fig.~\ref{fig:general}(a) and Fig.~\ref{fig:general}(b) respectively. Fig.~\ref{fig:general}(c) shows that in the absence of piezoelectric field the $s$ and the two lowest $d$ states are decoupled and the only anticrossing in this region appears between the $s$ state and the $d$ state with $M=0$ (e5 in Fig.~\ref{fig:smap}). The small splitting visible between the two lowest $d$ states is a numerical artefact caused by the discretization. The situation changes when the piezoelectric field is included. Then, the localization of one of the uncoupled $d$ states (e3 in Fig.~\ref{fig:smap}) is partially moved to the middle of the QD. In consequence, the symmetry changes and a coupling appears. However, the character of the second state (e4 in Fig.~\ref{fig:smap}) is unchanged, thus the second crossing still remains unsplitted.  
\begin{figure}[tb]
\begin{center}

\includegraphics[width=85mm]{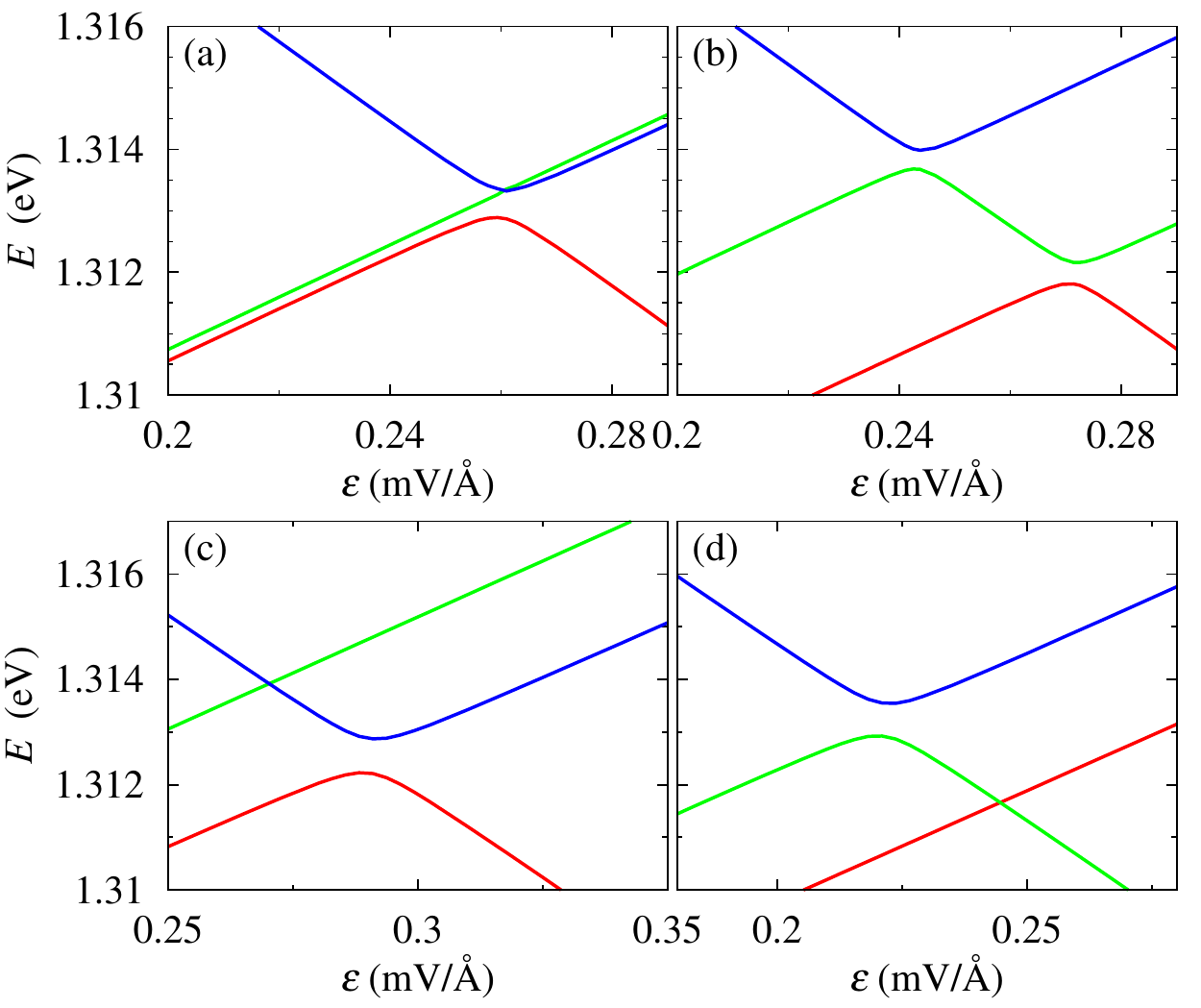}
\end{center}
\caption{\label{fig-shift}(Color online)  Energy branches of $s$ and $p$ states (a) without piezoelectric field for a shift of $x_\mathrm{s}=1.8$~nm, (b) with included piezoelectric field and a shift of $x_\mathrm{s}=1.8$~nm, (c)  with included piezoelectric field and $x_\mathrm{s}=\mathrm{y}_{s}=1.8$~nm, (d)  with included piezoelectric field  and  $x_\mathrm{s}=1.8$~nm, $y_\mathrm{s}=-1.8$~nm. } 
\end{figure}

In order to study the symmetry breaking effects in a DQD we displaced the lower dot in the plane perpendicular to the $z$ axis and we investigated the $s$-$p$ coupling. Both $p$ states are localized in the upper dot and the $s$ state is localized in the lower one. Fig.~\ref{fig-shift}(a) presents the energy branches for the interesting electric field range where the lower dot is shifted in the ($100$) direction by $x_{s}=1.8$~nm (that is $10$\% of the diameter of the lower dot) and the piezoelectric field is not taken into account.  The lowest $p$ state tends to be oriented along the direction of the displacement and the second one is perpendicular to it. As a result of the symmetry the $s$ state is coupled to the first $p$ state and uncoupled to the second one. The situation is different if the PZ field is included since this field forces alignment with respect to the ($110$) direction and this effect is much stronger than that resulting from the displacement (Fig.~\ref{fig-shift}(b)). Therefore, the $p$ states in upper dot are oriented along the ($110$) and ($1\bar{1}0$) direction respectively. In consequence, both the resonances between the $s$ and both $p$ states are opened and show a similar splitting in both cases.
However, if the QDs are displaced in the ($110$) direction then even in the presence of the PZ field, only one coupling is non-zero. As can be seen in Fig.~\ref{fig-shift}(c,d), a shift by $x_{\mathrm{s}}=y_{\mathrm{s}}=1.8$~nm and $x_\mathrm{s}=1.8$~nm $\mathrm{y}_{s}=-1.8$~nm opens only the first or the second $s$-$p$ resonance respectively. 

\begin{figure}[tb]
\begin{center}
\includegraphics[width=89mm]{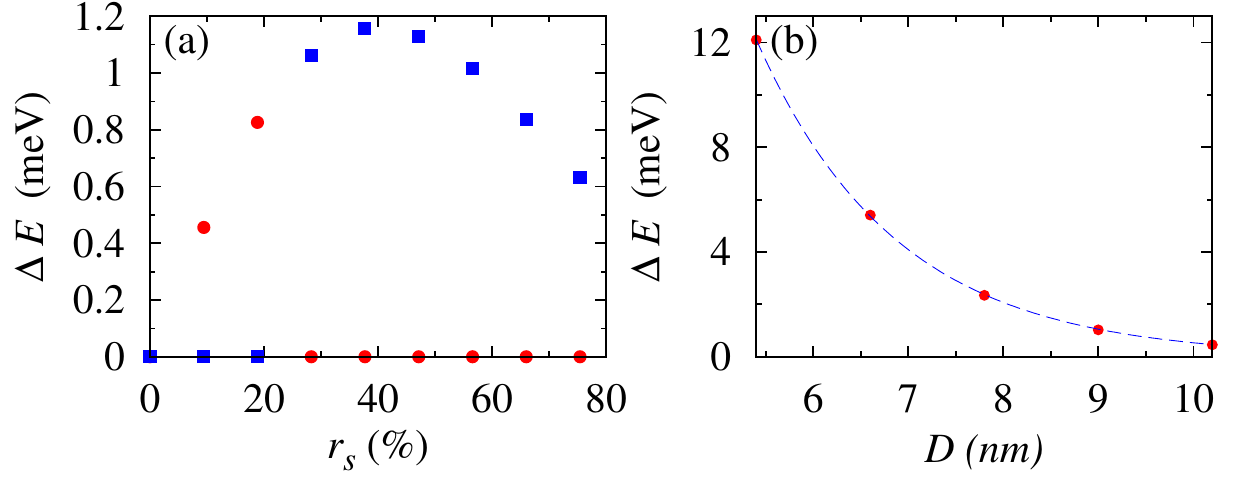}
\end{center}
\caption{\label{fig-ddep}(Color online) (a) The values of the energy splitting at the resonances between $s$ and lower $p$ state (red circles) and higher $p$ state (blue boxes) for $D=10.2$~nm as a function of the value of the relative displacement of the lower dot $r_{s}$,(b) The values of the energy splitting at the resonances between $s$ state and lower $p$ state for $x_\mathrm{s}=y_\mathrm{s}=1.8$~nm i.e., $r_{s}=14$\% as a function of $D$. The red points represents the simulation results and blue dashed line is an exponential fitting.} 
\end{figure}
We investigated the dependence of the $s$-$p$ coupling (as reflected by the width of the resonant splitting) on the value of the shift in the ($110$) direction. Fig.~\ref{fig-ddep}(a) shows the values of both resonant $s$-$p$ splittings as a function of the relative displacement $r_{s}=\sqrt{x^{2}_{s}+y^{2}_{s}}/{2 r_{1}}$  in the presence of the PZ field. 
In the case of a DQD with geometrical axial symmetry ($r_{s}=0$), the order of the $p$ states in the higher dot is opposite to the single QD case (e2,e3 in Fig.\ref{fig:smap}). It is caused by the influence of the PZ field from the lower dot\cite{usman11}. For a small shift in the ($110$) direction, the lower $p$ state is coupled and the second one remains decoupled. However, for shifts larger than about $25$~\%, the ordering of the $p$ states is reversed and the situation from a single QD is restored. In that case, the lower $p$ state is uncoupled and the higher one is coupled.
The dependence of the splitting on the value of the shift is determined by two processes: on the one hand, increasing of the displacement (in some range) enhances the $s$-$p$ coupling, but on the other hand the overlap between the wavefunctions decreases with the shift. As a result, the splitting has a maximum at a relative displacement near $40$\%.  
We investigated also the dependence of the $s-p$ splitting on the distance $D$ between the dots. Fig.~\ref{fig-ddep}(b) presents this splitting in the case of the constant shift value $x_\mathrm{s}=y_\mathrm{s}=1.8$~nm as a function of the distance $D$. As we can see, the dependence is nearly exponential. The splitting width 
($\Delta E$) is very well fitted by the formula $\ln (\Delta E/E_{0})=-\kappa D$, with parameters
$\kappa=0.678$~nm$^{-1}$ and $E_{0}=0.472$~eV (for the displacement $x_\mathrm{s}=y_\mathrm{s}=1.8$~nm).

\begin{figure}[tb]
\begin{center}
\includegraphics[width=89mm]{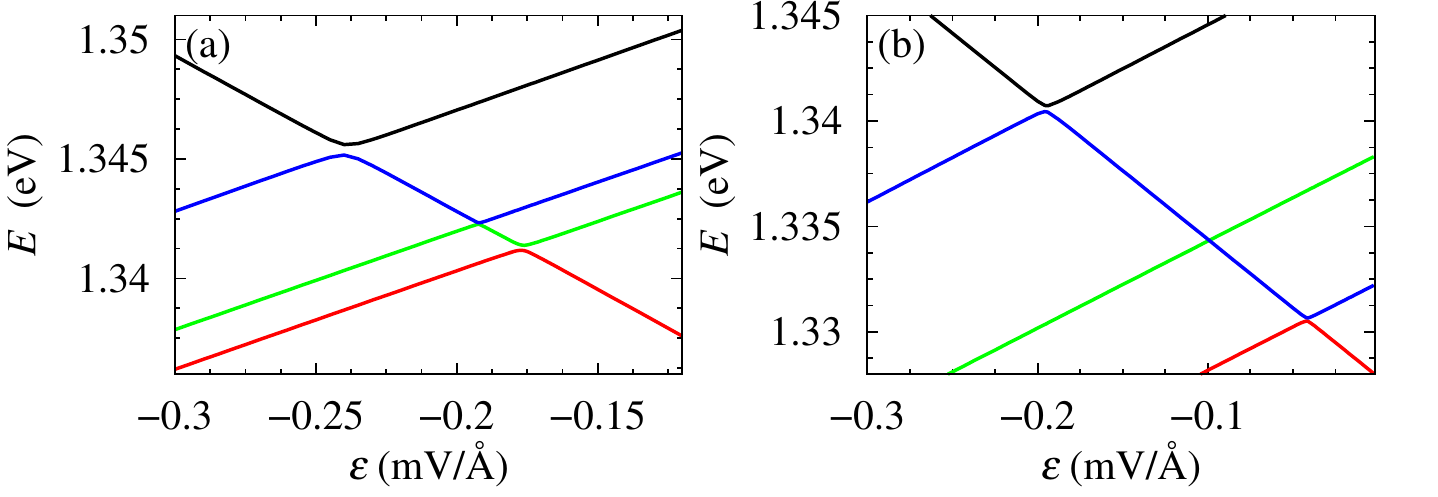}
\end{center}
\caption{\label{fig:elliptical}\textcolor{gray}({Color online) (a) Energy branches as a function of the electric field  in the region of the $s$-$d$ resonances at $x_\mathrm{s}=1.8$~nm with $y_\mathrm{s}=0$. 
(b) Energy branches as a function of electric field in the region of the $s$-$d$ resonances in the case of elliptical dots.} }
\end{figure}

We also investigated the influence of symmetry breaking on the $s$-$d$ coupling. As presented in Fig.~\ref{fig:general}(d), the coupling between the $s$ and the two $d$ states can appear even if the geometrical symmetry is conserved and is a consequence of the PZ field. Here, we investigate the effects of symmetry breaking in two cases: the shift along the ($100$) direction and the situation where both dots are elliptical. In the case of the shift (Fig.~\ref{fig:elliptical}(a)) by $x_\mathrm{s}=1.8$~nm with $y_\mathrm{s}=0$, a mixing between the $s$ state and the second $d$ state becomes possible. However, this coupling is weak and only a small splitting in the resonance appears (of the order of $66\mu$eV).  Also for the first and third $d$ state, the effect is relatively small and we obtain splitting values comparable to those resulting only from the PZ field. The reason is that a shift in the ($100$) direction conserves the mirror symmetry in ($010$) direction which is also important from the point of view of the coupling. 

 In the next step, we consider both dots to have an elliptical shape with the major to minor axis ratio of $1.1$ where the elongation is in the ($110$) direction. Due to the symmetry reasons, ellipticity does not lead to $s$-$p$ mixing. Although this mixing appears if the $SO$ coupling is included\cite{daniels13}, its value is small (about $100\mu$eV at $D=10$nm). The results for elliptical dots in the region of the $s$-$d$ resonances are shown in  Fig.~\ref{fig:elliptical}(b). From the qualitative point of view, an elongation in ($110$) direction does not change the situation from Fig.~\ref{fig:general}(d), that is, $s$ is coupled only to the first and third $d$ state. However, this leads in particular to a reduction of the width of the resonance between the $s$ and the lowest $d$ state.

\section{Conclusions} 
\label{sec:concl}
In summary, we have studied the effects of coupling between the electron states from different subshells ($s,p,d$) localized in different dots in a DQD structure taking into account the orientation of the system geometry with respect to the crystallographic axes. We have shown that breaking of the geometrical axial symmetry by a relative off-axial shift of the dots can lead to significant $s$-$p$ mixing.  We have found out, that in the presence of the piezoelectric field the direction of the shift plays an important role in the electronic structure. We have also shown that $s$-$p$ resonances are much more sensitive to the geometrical symmetry breaking than $s$-$d$ resonances. We have studied the influence of dot ellipticity on the $s$-$d$ resonances, and we have shown that those effects gives only quantitative contribution to the effect resulting from the piezoelectric field.

\acknowledgments

We are grateful to Janusz Andrzejewski and Jonas Daniels for inspiring discussion as well as to Thomas Kendziorczyk for many useful numerical clues.
This work was supported by the Foundation for Polish Science under the TEAM programme,
co-financed by the European Regional Development Fund. K.G. acknowledges support from the German Academic Exchange Service (DAAD). 

\appendix

\section{Piezoelectric field} 
\label{sec:app1}

In order to calculate the potential generated by the piezoelectricity, we found the polarization $\bm{P}=\bm{P_{1}}+\bm{P_{2}}$ up to second order in strain tensor. In the case of zincblende structure growth in the ($001$) direction, it takes the form \cite{bester06b,schultz11,schultz12}
\begin{displaymath}
\bm{P_{1}}  =   e_{14} \begin{pmatrix} \epsilon_{yz} \\  \epsilon_{xz} \\  \epsilon_{xy} \end{pmatrix} ,
\end{displaymath}
\begin{align*}
\bm{P_{2}}  = &  2B_{114} \begin{pmatrix} \epsilon_{xx} \epsilon_{yz} \\ \epsilon_{yy} \epsilon_{xz} \\ \epsilon_{zz} \epsilon_{xy} \end{pmatrix} + 
2B_{124} \begin{pmatrix} (\epsilon_{yy} + \epsilon_{zz} ) \epsilon_{yz} \\ (\epsilon_{xx} + \epsilon_{zz} ) \epsilon_{xz} \\ (\epsilon_{xx} + \epsilon_{yy} ) \epsilon_{xy} \end{pmatrix} \\ & + 2B_{156} \begin{pmatrix} \epsilon_{xz} \epsilon_{xy} \\ \epsilon_{yz} \epsilon_{xy} \\ \epsilon_{yz} \epsilon_{xz} \end{pmatrix},
\end{align*}
where $e_{14}$ and $B_{114},B_{124},B_{156}$ are linear and quadratic polarization parameters respectively. Then, the piezoelectricity-induced charge is calculated from $\rho_{\mathrm{piezo}}=-\nabla \cdot \bm{P}$. Finally, the piezoelectric potential $V_{p}$ is found from the solution of the Poisson-like equation
\begin{displaymath}
\rho_{\mathrm{piezo}}  =    \varepsilon_{0} \nabla [  \varepsilon_{S}(\rr) \nabla V_{p} ] ,
\end{displaymath}
where $\varepsilon_{S}(\rr)$ is the position-dependent static dielectric constant.

\section{Calulation details} 
\label{sec:app2}

We have performed the calculation of the strain tensor as well as the electron states. We have used a non-uniform grid ($160$~x~$160$~x~$160$) with mesh size nearly half of the InAs lattice constant ($0.3$~nm) inside the QDs and with size linearly increasing outside the QDs. In order to calculate the displacement field and the piezoelectric field, we have solved numerically a linear set of equations using the GMRES method combined with the ILU preconditioner with the LIS library\cite{lis}. The electron states are found using Jacobi-Davison method (which allows us to obtain the eigenvalues from the middle of the energy spectrum due to the spectral transformation) in the SLEPC library\cite{slepc} combined with the PETSC library\cite{petsc}.  Second order derivatives have been discretized according to\cite{tan90}
\begin{align*}
\frac{d}{dx_{k}} \left ( A\frac{d}{dx_{k}}B \right ) = & \frac{B_{i+1}-B_i{}}{h_{i}(h_{i}+h_{i-1})} \left ( A_{i+1} + A_{i} \right ) \\
& + \frac{B_{i-1} - B_{i}}{h_{i}(h_{i}+h_{i-1})} \left ( A_{i-1} + A_{i} \right ),
\end{align*}
where $h_{i}$ is a position-dependent mesh size. Such discretization leads to an asymmetric matrix in the eigenvalue problem. In order to restore symmetrization, we applied an appropriate transformation as described in Ref.~\onlinecite{tan90}. 

\bibliographystyle{prsty}
\bibliography{abbr,quantum}
\end{document}